\documentclass[conference]{IEEEtran}
\ifCLASSINFOpdf
\else
\fi
\usepackage{algorithm}
\usepackage{algorithmic}
\usepackage{graphicx}

\begin{document}
\title{Adaptive Scheduling in Real-Time Systems Through Period Adjustment}
\author{\IEEEauthorblockN{Shri Prakash Dwivedi}
\IEEEauthorblockA{Department of Computer Science and Automation\\
Indian Institute of Science\\
Bangalore, India\\
Email: shriprakashdwivedi@gmail.com}}
\maketitle

\begin{abstract}
Real-time system technology traditionally developed for safety critical systems, has now been extended to support multimedia systems and virtual reality. A large number of real-time application, related to multimedia and adaptive control system, require more flexibility than classical real-time theory usually permits. This paper proposes an efficient adaptive scheduling framework in real-time systems based on period adjustment. Under this model periodic tasks can change their execution rates based on their importance values to keep the system underloaded. We propose Period\_Adjust algorithm, which consider the tasks whose periods are bounded as well as the tasks whose periods are not bounded.
\end{abstract}
\IEEEpeerreviewmaketitle

\section{Introduction}
The real-time scheduling paradigms, both static such as rate
monotonic scheduling [13], and dynamic such as earliest deadline first scheduling, do not fit well the requirements of advanced real-time applications in dynamic environments. Real-time systems are being increasingly designed for complex systems. For these applications, it is sometime impractical or impossible to provide  static guarantees to real-time computation. These motivations have led to the emergence of the adaptation and overload management as a major research issue in real-time systems. \\
\indent An overview of prior art in overload management and adaptive scheduling techniques for real-time systems is given in Lu et al. [14]. Mechanism for detecting and handling timing errors including overloads are discussed in Stewart and Khosla [20], with emphasis on a specific application-oriented operating systems. An interesting technique for overload management in hard real-time control applications is described in Ramanathan et al. [17]. The author presents a scheduling policy deterministically guaranteeing $m$ out of any $k$ periodic task activations, along with a methodology able to minimize the effects of missed control-law updates. This work provides a solid foundation to graceful degradation policies of periodic real-time tasks. However, unless the overload duration is very short, the application could be significantly impaired by the loss of periodic execution for a number of real-time tasks.
Dynamic Window Constrained Scheduling algorithm is similar except that the window $k$ is fixed. Mok et al. [16] modified Dynamic Window Constraint Scheduling, which is primarily deadline based by using the  concept of Pfairness to improve the success rate for tasks with unit size execution time. Other frameworks such as the imprecise computation model  and reward based model can be applied in the situation where quality of service is proportional to the amount of workload completed.\\
\indent The need for adaptive management of the Quality of Service has been widely recognized in the domain of the distributed multimedia systems.  A graceful degradation of the communication subsystem is obtained in Abdelzaher and Shin [1] by means of QoS contracts specifying degraded acceptable QoS levels. Significant research has also been devoted to schedulers providing some degree of adaptation to cope with the dynamic overload environment. The need for scheduling systems providing real-time guarantee to a subset of tasks within a general operating system has been emphasized in the Stankovic et al. [19]. In Lu et al. [14] the authors assume a flexibility in timing requirements. To address the dynamics of the environment, they  proposed a modified EDF adaptive scheduling framework based on feedback
control methods and use feedback control loops to maintain a satisfactory deadline miss ratio when task execution times change.\\
\indent Many real-time task models have been proposed to extend timing requirements beyond the hard and soft deadlines based on the observation that jobs can be dropped without severely affecting performance [4]. Despite the success of some models in alleviating overload situation, it is sometime more suitable to execute jobs less often instead of dropping them or allocating fewer cycles. The work in Kuo et al. [12] is among the first to address this type of requirement. Load-adjustable algorithms and value-based policies are the main techniques proposed for graceful recovery from overload. A load adjustment mechanism is proposed in [12] in order to handle periodic processes with varying temporal parameters. The aim of this work is to determine feasible time parameter configurations (execution time $C$ and period $T$) and thus modify the real-time computation for collections of tasks. The configuration selection problem is solved by a harmonic approach achieving the maximum exploitation of the computational resources under any time parameter configuration. While appealing, this approach does not lend itself to many real-time systems, where execution times, in spite of their variability, cannot be set or chosen by the designer.\\
\indent In [18] Seto et al. considered the problem of finding a feasible set of task period as a non-linear programming problem, which seeks to optimize specific form of control performance measure. Cervin et al. used optimization theory to solve the period selection problem online by adaptively adjusting task periods with focus on optimizing specific control performance [9]. Baruah et al. [2] proposed a scheduling algorithm maximizing the effective processor during overload, given a minimum slack factor for all tasks.\\
\indent Buttazo et al. [5] proposed a flexible framework known as  elastic task model, where deadline misses are avoided by increasing task  periods until some desirable utilization is achieved. The work in [14] extends the basic elastic task model to handle cases where the computation time is unknown. In elastic task model [6],[7], periodic computations are modeled as springs with given elastic coefficients and minimum lengths. Requested variations in task execution rates or overload conditions are managed by changing the rates based on the spring's elastic coefficients.
Generalized elastic scheduling proposed by Chantem et al. [10],[11], by generalizing elastic scheduling approach. Although the Elastic model is nice but it does not consider the cases where the task periods of soft real-time systems may be unbounded or loosely bounded. We develop in this paper an efficient adaptive scheduling scheme in real-time systems through period adjustment, which consider the tasks having bounded as well as unbounded periods.\\
\indent This paper is organized as follows. Section 2 describes problem definition and motivation. Section 3 presents our proposed task model and the Period\_Adjust algorithm and its features. In section 4, we present the experimental results. Finally, section 5 contains conclusion.
\section{Problem Definition and Motivation}
Many models have been proposed in real-time scheduling theory to deal with adaptive scheduling and overload management. Some of the proposed models are based on the observation that less important jobs can be dropped without severely affecting performance. But dropping of jobs may not always be the best option, because it is sometime more suitable to execute the jobs less often instead of dropping them even if they are less important. Elastic task model [6] uses flexible framework but it do not consider the case where some of the soft real-time task may be loosely bounded or unbounded. We propose a novel scheduling framework based on period adjustment. Our algorithm considers the tasks whose periods are tightly bounded as well as the tasks whose periods are loosely bounded. We feel that this is more general model and this model performs nicely even when all tasks are bounded.\\
\indent
Many soft real-time applications require the execution of periodic
activities, whose rate can usually be defined within a certain range. The higher the frequency, the better the performance. Depending on the application domain, some tasks are rigidly imposed by the environment
whereas other activities can be more flexible, producing significant results when their rates are within a certain range. For example, in multimedia systems the activities such as voice sampling, image acquisition, data compression, and video playing are performed periodically, but their execution rates are not so rigid. Depending on the requested quality of service, tasks may increase or decrease their execution rate to accommodate the requirements of other concurrent activities. However this period range may be flexible also. Suppose a soft real-time task has period range $(a,b)$, then in some application it may be possible to increase few time units above $b$ and decrease few time units below $a$, if by doing so system become schedulable. It is sometime counter intuitive that a soft real-time application which is schedulable in range $(a,b)$ can not be schedulable for the range  $(a,(b+1))$ or alike. There are many flexible applications in multimedia and control applications in which we may be able to vary few time units across bound (upper or lower) without severely affecting the performance. We feel that there should be a general scheduling framework which can consider the flexible applications whose periods are unbounded alongwith the bounded one.

\section{Proposed Work}
\subsection{Task Model}
We consider the system where each task $\tau_i$ is periodic and is characterized by the following tuple: ($C_i$,$T_{i_0}$,$T_{i_{\min}}$,$T_{i_{\max}}$,$w_i$) for $i=1,\ldots N$. Where $N$ is the number of tasks in the system, $C_i$ is the worst case execution time  and $T_{i_0}$ is the initial period of $\tau_i$. $T_{i_{\min}}$ denotes the minimum possible period of $\tau_i$ as specified by the application, and $T_{i_{\max}}$ represents the maximum period beyond which the system performance is no longer acceptable. The weighting factor $w_i$, represents importance of task $\tau_i$, to changing it's period in face of changes. The longer the weighting factor of a task, the more will be it's contribution towards the overall utilization. Given a task set $\Gamma$, tasks are arranged in a nondecreasing order of deadline.\\
\indent Each task $\tau_i$ in task set $\Gamma$ is divided in to two parts. $\Gamma_h$ for hard real-time tasks and $\Gamma_s$ for soft real-time tasks such that $\Gamma = \left\{ \Gamma_h \cup \Gamma_s \right\}$. $N$ is the total number of tasks in the systems. $n_h$ is the number of hard real-time tasks such that $(n_h=\left| \Gamma_h \right|)$. $n_s$ is the number of soft real-time tasks such that $(n_s=\left| \Gamma_s \right|)$. $w_i$ is the weighting factor or importance value of each soft real-time tasks in $\Gamma_s$. $w_i$'s for soft real-time tasks are arranged in such a way that $\sum_{i =1}^{n_s} w_i=1$, in other words $w_i$ represents fractional importance value or percentage of importance value of each soft real-time tasks towards the whole system performance. Furthermore a task in $\Gamma_s$ may belongs to $\Gamma_{s\_{bound}}$ or $\Gamma_{sp}$ or $\Gamma_{s\_{unbound}}$, i.e. ${\Gamma_s = \{\Gamma_{s\_{bound}} \cup \Gamma_{sp} \cup \Gamma_{s\_{unbound}}\}}$ where $\Gamma_{s\_{bound}}$ consists of those soft real-time tasks for which an upper bound or lower bound or both are imposed on tasks periods prior to execution or during execution, and $\Gamma_{sp}$ consists of those soft real-time tasks which have fixed periods or which requests for fixed periods during run time, whereas task set $\Gamma_{s\_{unbound}}$ consists of those tasks which are unbounded. However as a matter of fact period $T_i$ can not be less than worst case computation time $C_i$ of a task. Our scheduling algorithm emphasize such soft real-time application which have more number of tasks in $\Gamma_{s\_{unbound}}$.\\
\indent In this task model, all the tasks $\tau_i$, which does not belongs to $\Gamma_{s\_{bound}}$ can have $T_{i_{min}}$ or $T_{i_{max}}$ equal to $\Phi$, which means that they are unbounded. For each $(\tau_i \in \Gamma_h)$, $w_i=h_{rt}=1$ which means that all hard real-time tasks must execute provided they are schedulable. $T_i$  denotes the actual period of task $\tau_i$, which is constrained to be in the range [$T_{i_{\min}}$, $T_{i_{\max}}$] for the case $(\tau_i \in \Gamma_{s\_{bound}})$, whereas $C_i$ denotes actual execution time considered to be known a priori. In the case of tasks with variable computation times, $C_i$ will denote the actual worst case execution time. Any period variation is always subject to an utilization guarantee and is accepted only if there exists a feasible schedule such that tasks are scheduled by earliest deadline first algorithm. Hence if $\sum (C_i / T_{i_0}) \leq 1$, all tasks can be created at the minimum period $T_{i_0}$, otherwise the algorithm is used to adapt the task's period to $T_i$ such that $\sum (C_i / T_{i}) = U_d \leq 1$, where $C_i$ is the actual online execution estimate and $U_d$ is some desired utilization factor.
System designer can set $w_i$ statically or dynamically depending upon requirements. In static method, all soft real-time tasks are assigned $w_i$'s prior to start of the task execution and these $w_i$'s remains fixed up to the end of the task completions. In dynamic method, assignment of $w_i$ is event based i.e., weighting factor $w_i$ may be reassigned during the occurrence of any event such as, a new task arrival or completion of a task.

\subsection{Period\_Adjust Algorithm}
We propose a new scheduling framework namely Period\_Adjust algorithm
which accepts set of tasks $\Gamma$ and desired utilization $U_d$ and return set of periods for soft real-time tasks so as to maximize quality of service. We may set $U_d$ equal to the maximum schedulable utilization of individual scheduling algorithm. We can set $U_d = 1$ for dynamic scheduling algorithm like EDF, or we can set $U_d (n) = n (2^{1 / n} - 1)$ for the static scheduling RM algorithm, where $n$ is the number of independent, preemptable periodic tasks with relative deadline equal to their respective periods. In this algorithm we assume that deadline is equal to the period. We also assume that the execution time $C_i$ of all the tasks is given prior alongwith the periods of hard real-time tasks. The total task set $\Gamma$ is divided in two groups, namely the set of hard real-time tasks $\Gamma_h$, and the set of soft real-time tasks $\Gamma_s$. Furthur the set of soft real-time tasks may consists of $\Gamma_{sp}$, in which soft real-time task request for fixed period, $\Gamma_{s\_{bound}}$ in which tasks are bounded by maximum and minimum periods.\\
\indent Our Period\_Adjust algorithm works as follows: The first for loop computes the utilization of hard real-time tasks, then algorithm computes the summation of all utilization of task set $\Gamma_h$ to check for its feasibility. In the second for loop it computes the utilization of those tasks which request for period change, if there is no such task $U_{sp}$ is set to zero, after that it again checks for the feasibility of the schedulable utilization. The third for loop computes the tasks periods of all soft real time tasks in accordance with their weighting factor or importance value. Next the algorithm  checks whether the periods of unbounded tasks are less than their computation time. If period is less than computation time, it replaces period by computation time. Finally it checks whether these periods exceeds their bounds for the bounded tasks, if this is the case it replaces periods with their bounds. 
\begin{algorithm}
\caption{\bf :  Period\_Adjust$(\Gamma, U_d)$}
\begin{algorithmic}
\FOR { each $(\tau_i \in \Gamma_h)$ } 
 {
\STATE $U_i = \frac{C_i}{T_i}_{}$ \\
 }
\ENDFOR
\STATE $U_h = \sum U_i$
\STATE $U_s = U_d - U_h$
\IF {$(U_s \leq 0)$}
\RETURN $infeasible$
\ENDIF
\FOR {each$(\tau_i \in \Gamma_{sp})$ }

\STATE $U_i = \frac{C_i}{T_{i_{sp}}}_{}$

\ENDFOR
\STATE $U_{sp} = \sum U_i$
\STATE $U_s = U_d - U_h - U_{sp}$
\IF {$(U_s \leq 0)$}
\RETURN $infeasible$
\ENDIF
\FOR {each$(\tau_i \in (\Gamma_s-\Gamma_{sp})$ }
\STATE $T_i = \frac{C_i}{\Big(w_i + \frac{\sum w_{{sp}_i}}{N-n_h-n_{sp}}\Big)(U_d - U_h - U_{sp})}$
\RETURN $T_i$
\ENDFOR
\FOR {each$(\tau_i \in (\Gamma_s-\Gamma_{sp}-\Gamma_{s\_bound}))$ }
\IF {$(T_i < C_i)$}
\STATE $T_i=C_i$
\ENDIF
\ENDFOR
\STATE $mod=0$
\FOR {each$(\tau_i \in \Gamma_{s\_bound})$}
 \IF {$(T_i < T_{i_{\min}})$}
 \STATE $T_i = T_{i_{\min}}$
 \ELSE
      \IF {$(T_i > T_{i_{\max}})$}
\STATE $T_i =  T_{i_{\max}}$
\STATE $\Gamma_{s\_bound} = \Gamma_{s\_bound} - \tau_i$
\STATE $\Gamma_{sp} = \Gamma_{sp} + \tau_i$
\STATE $mod=1$
  \ENDIF \\
\ENDIF
\ENDFOR
\IF {$(mod==1)$}
\RETURN Period\_Adjust$(\Gamma, U_d)$
\ELSE
\RETURN $feasible$
\ENDIF
\end{algorithmic}

\end{algorithm}

If computed period $T_i$ for a bounded task is less than the minimum period $T_{i_{min}}$, we can simply replace $T_i$ by $T_{i_{min}}$, because increasing the period leads to less overall utilization. However, if the computed period $T_i$ is greater than the maximum period $T_{i_{max}}$, we can not simply replace $T_i$ by $T_{i_{max}}$, because decreasing the period leads to increased utilization, which may exceeds the schedulable utilization. Therefore corresponding task is removed from bounded task set $\Gamma_{s\_{bound}}$ to fixed period task set $\Gamma_{sp}$ and Period\_Adjust algorithm is re-invoked. In this algorithm we assume that in soft real-time application there are many cases where either no bounds are available or no bounds are required for soft real-time tasks.

\section{Experimental Results}
In this section we present the experimental results performed on our task model. We consider period selection with deadlines equal to periods. In all the following tables here onwards periods $(T_{i_0}, T_{i_{\min}}, T_{i_{\max}})$ and computation times $(C_i)$ are expressed in milliseconds(ms).
\begin{table}[ht]
\renewcommand{\arraystretch}{1.3}
\caption{Task Set Parameters}
\label{table 1}
\begin{center}
\begin{tabular}{|c| c| c| c| c| c|} 
\hline
Task & $C_i$ & $T_{i_{0}}$ & $T_{i_{min}}$ &  $T_{i_{max}}$ & $w_i$  \\ [0.5ex] \hline\hline 
$\tau_1$ & 18 & 100 & 50 & 150 & 0.30 \\ \hline
$\tau_2$ & 18 & 100 & 50 & 150 & 0.30 \\ \hline
$\tau_3$ & 18 & 100 & 50 & 150 & 0.18 \\ \hline
$\tau_4$ & 18 & 100 & 50 & 150 & 0.12 \\ \hline
$\tau_5$ & 18 & 100 & 50 & 150 & 0.10 \\ [1ex] \hline
\end{tabular} 
\end{center}
\end{table} 

\indent To execute the Period\_Adjust algorithm, we first use the task set parameters given in Table 1. In this experiment, all tasks starts at time 0 with an initial period of 100 time units and the task set is schedulable under EDF. Here the required maximum utilization of the overall system is $\frac{18}{50} + \frac{18}{50} + \frac{18}{50} + \frac{18}{50} + \frac{18}{50} = 1.8$, whereas the required minimum utilization of the overall system is $\frac{18}{150} + \frac{18}{150} + \frac{18}{150} + \frac{18}{150} + \frac{18}{150} = 0.6$. Since the current utilization is $\frac{18}{100} + \frac{18}{100} + \frac{18}{100} + \frac{18}{100} + \frac{18}{100} = 0.90$, the task set is schedulable under EDF. Assume that, at the 10sec, $\tau_{1_{}}$ needs to reduce its period to 50 time units, due to some changes in system dynamics not experienced by other tasks. Since the new required utilization of the system is $\frac{18}{50}_{} + \frac{18}{100} + \frac{18}{100} + \frac{18}{100} + \frac{18}{100} = 1.08$. which is greater than 1, and therefore as such it is not schedulable under EDF. We can observe that the required minimum utilization of the system is $\frac{18}{50} + \frac{18}{150} + \frac{18}{150} + \frac{18}{150} + \frac{18}{150} = 0.84$, which is less than 1. Therefore to allow for $\tau_1$ to change its period, the period of tasks $\tau_2$, $\tau_3$, $\tau_4$ and $\tau_5$ must increase for the system to remain schedulable. At time 20sec, $\tau_1$ goes back to its original period state. Fig. 1 shows the cumulative number of executed instances for each task as its period changes over time. When we execute Period\_Adjust algorithm on the above task sets, it will return the feasible set of task periods$(T_1 = 50, T_2 = 80, T_3 = 110, T_4 = 138, T_5 = 150)$.

\indent Now we consider the same task set parameters with some change. Here we assume that soft real-time tasks $\tau_4$ and $\tau_5$ are not bounded, i.e. although the preferable maximum period is 150, some flexibility is provided by the application to increase or decrease the bound. In this case assume that at 10sec \ $\tau_{1_{}}$ needs to reduce the its period to 50 time units and $\tau_{2_{}}$ needs to reduce the its period to 60 time units, as shown in Table 2.
\begin{table}[ht]
\renewcommand{\arraystretch}{1.3}
\caption{Task Set Parameters}
\label{table 2}
\begin{center}
\begin{tabular}{|c| c| c| c| c| c|} 
\hline
Task & $C_i$ & $T_{i_{0}}$ & $T_{i_{min}}$ &  $T_{i_{max}}$ & $w_i$ \\ [0.5ex] \hline\hline 
$\tau_1$ & 18 & 50 & 50 & 150 & 0.30 \\ \hline
$\tau_2$ & 18 & 60 & 50 & 150 & 0.30 \\ \hline
$\tau_3$ & 18 & 100 & 50 & 150 & 0.18 \\ \hline
$\tau_4$ & 18 & 100 & $\Phi$ & $\Phi$ & 0.12 \\ \hline
$\tau_5$ & 18 & 100 & $\Phi$ & $\Phi$ & 0.10 \\ [1ex] \hline
\end{tabular} 
\end{center}
\end{table} 

For these task set parameters Task\_compress algorithm [5] is infeasible, whereas Period\_Adjust algorithm is feasible. In fact when we execute the Period\_Adjust algorithm on the above task sets, the corresponding periods obtained for the tasks are shown in Fig. 2 $(T_1 = 50, T_2 = 60, T_3 = 147,  T_4 = 155, T_5 = 175)$.

\indent Now, we consider the task set parameters given in Table 3 for the case of admission control policy during dynamic task activation.
\begin{table}[ht]
\renewcommand{\arraystretch}{1.3}
\caption{Task Set Parameters}
\label{table 3}
\begin{center}
\begin{tabular}{|c| c| c| c| c| c| } 
\hline
Task & $C_i$ & $T_{i_{0}}$ & $T_{i_{min}}$ &  $T_{i_{max}}$ & $w_i$  \\ [0.5ex] \hline\hline 
$\tau_1$ & 30 & 100 & 50 & 350 & 0.20 \\ \hline
$\tau_2$ & 50 & 200 & 50 & 350 & 0.20 \\ \hline
$\tau_3$ & 70 & 300 & 50 & 350 & 0.20 \\ \hline
$\tau_4$ & 10 & 100 & 50 & 350 & 0.20 \\ \hline
$\tau_5$ & 10 & 70 & 50 & 350 & 0.20 \\ [1ex] \hline
\end{tabular}
\end{center}
\end{table} 
In this experiment $\tau_1$, $\tau_2$ and $\tau_3$ starts at time 0. They have the current utilization $\frac{30}{100} + \frac{50}{200} +
\frac{70}{300} = 0.78$ and therefore schedulable by EDF. At time 10sec two tasks $\tau_4$ and $\tau_5$ arrives which makes the total utilization $\frac{30}{100} + \frac{50}{200} + \frac{70}{300} + \frac{10}{100} + \frac{10}{70} = 1.03$. In order to allow the tasks $\tau_4$ and $\tau_5$ for execution, the tasks  $\tau_1$, $\tau_2$ and $\tau_3$ can increase their period. Since both tasks $\tau_4$  and $\tau_5$ are of 10\,sec duration, after 20\,sec  tasks $\tau_1$, $\tau_2$ and $\tau_3$ returns to their previous periods, as shown in the Fig. 3(Dynamic task activation).
Now we consider the above task set parameters with some modification. In this  case $\tau_4$ and $\tau_5$ arrives at 10\,sec having the computation times 30\,ms and 20\,ms respectively as shown in Table 4. Here task $\tau_3$ is loosely bounded (period of task $\tau_3$ should be preferably between 50 and 350 but not necessarily). In this case total utilization is $U = \frac{30}{100} + \frac{50}{200} + \frac{70}{300} + \frac{30}{100} + \frac{20}{70} = 1.37.$ Obviously task sets are not schedulable. Task set parameters alongwith importance values are given in the follwing table.
\begin{table}[ht]
\renewcommand{\arraystretch}{1.3}
\caption{Task Set Parameters}
\label{table 4}
\begin{center}
\begin{tabular}{|c| c| c| c| c| c| } 
\hline
Task & $C_i$ & $T_{i_{0}}$ & $T_{i_{min}}$ &  $T_{i_{max}}$ & $w_i$  \\ [0.5ex] \hline\hline 
$\tau_1$ & 30 & 100 & 50 & 350 & 0.20 \\ \hline
$\tau_2$ & 50 & 200 & 50 & 350 & 0.20 \\ \hline
$\tau_3$ & 70 & 300 & $\Phi$ & $\Phi$ & 0.20 \\ \hline
$\tau_4$ & 30 & 100 & 50 & 350 & 0.20 \\ \hline
$\tau_5$ & 20 & 70 & 50 & 350 & 0.20 \\ [1ex] \hline
\end{tabular}
\end{center}
\end{table} 
In this case also Task\_compress algorithm is infeasible. While
Period\_Adjust algorithm is feasible. On execution periods returned by the Period\_Adjust algorithm are $(T_1 = 150, T_2 = 250, T_3 = 355, T_4 = 150,  T_5 = 200)$.

\begin{figure*}[ht]
\begin{center}
\includegraphics[width=12cm, height=9cm]{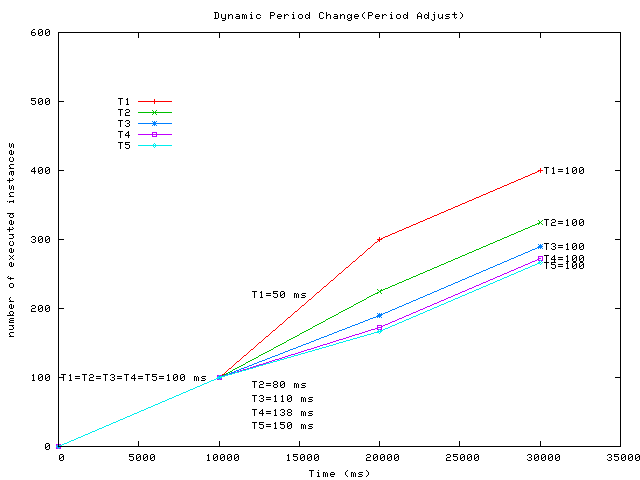}
\end{center}
\caption{Dynamic period change using Period\_Adjust }
\label{graph4}
\end{figure*}

\begin{figure*}[ht]
\begin{center}
\includegraphics[width=12cm, height=9cm]{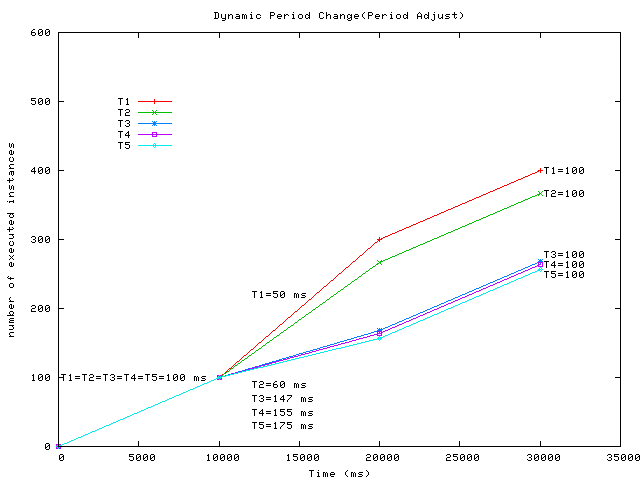}
\end{center}
\caption{Dynamic period change which is feasible by Period\_Adjust only }
\label{graph5}
\end{figure*}

\begin{figure*}[ht]
\begin{center}
\includegraphics[width=12cm, height=9cm]{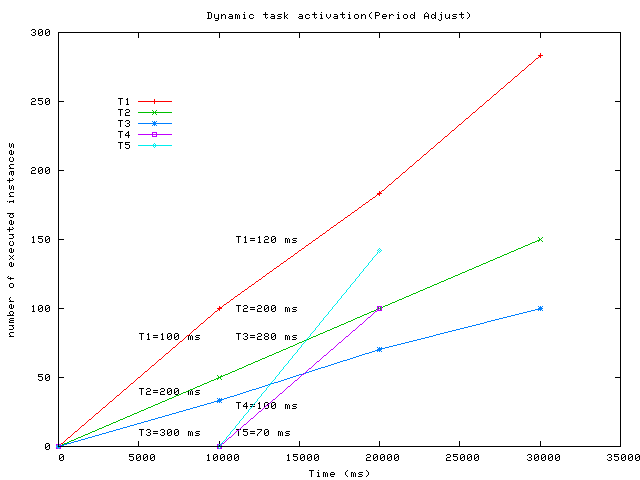}
\end{center}
\caption{Dynamic Task Activation using Period\_Adjust }
\label{graph6}
\end{figure*}

\indent For the comparison purpose, here we  use the task set parameters in [7], and we show that Period\_Adjust works nicely in these cases also.

\begin{table}[ht]
\renewcommand{\arraystretch}{1.3}
\caption{Task Set Parameters}
\label{table 5}
\begin{center}
\begin{tabular}{|c| c| c| c| c| c| c|} 
\hline
Task & $C_i$ & $T_{i_{0}}$ & $T_{i_{min}}$ &  $T_{i_{max}}$ & $E_i$ & $w_i$ \\ [0.5ex] \hline\hline 
$\tau_1$ & 24 & 100 & 30 & 500 & 1 & 0.30 \\ \hline
$\tau_2$ & 24 & 100 & 30 & 500 & 1 & 0.30 \\ \hline
$\tau_3$ & 24 & 100 & 30 & 500 & 1.5 & 0.25 \\ \hline
$\tau_4$ & 24 & 100 & 30 & 500 & 2 & 0.15 \\ [1ex] \hline
\end{tabular} 
\end{center}
\end{table} 

Task set parameters are shown in Table 5. In this experiment four periodic tasks are created at time $t=0$. All the tasks start executing at their initial period, at $t = 10$\,sec $\tau_1$ decreases its period from 100\,ms to 33\,ms. At $t = 20$\,ms $\tau_1$ returns to its initial period. The result of the application of Period\_Adjust algorithm and Task\_compress algorithm on the above task sets is shown in the Fig. 4. It shows the actual number of instances executed by each task as a function of time.
Next experiment consider the case of admission control policy during
dynamic task activation (Table 6). Three tasks starts executing at the time $t = 0$ at their initial period. An other task $\tau_4$ arrives at time $t = 10$\,sec. Since tasks are not schedulable when $\tau_4$ is started, Period\_Adjust algorithm is invoked which increases the periods of other tasks to make the request of task $\tau_4$ fulfilled.

\begin{table}[ht]
\renewcommand{\arraystretch}{1.3}
\caption{Task Set Parameters}
\label{table 6}
\begin{center}
\begin{tabular}{|c| c| c| c| c| c| c|} 
\hline
Task & $C_i$ & $T_{i_{0}}$ & $T_{i_{min}}$ &  $T_{i_{max}}$ & $E_i$ & $w_i$ \\ [0.5ex] \hline\hline 
$\tau_1$ & 30 & 100 & 30 & 500 & 1 & 0.25 \\ \hline
$\tau_2$ & 60 & 200 & 30 & 500 & 1 & 0.25 \\ \hline
$\tau_3$ & 90 & 300 & 30 & 500 & 1 & 0.25 \\ \hline
$\tau_4$ & 24 & 50 & 30 & 500 & 1 & 0.25 \\ [1ex] \hline
\end{tabular} 
\end{center}
\end{table} 

Fig. 5 shows the actual number of instances executed by each task as a
function of time during the execution of the Period\_Adjust algorithm and Task\_compress Algorithm.

\begin{figure*}[ht]
\begin{center}
\includegraphics[width=12cm, height=9cm]{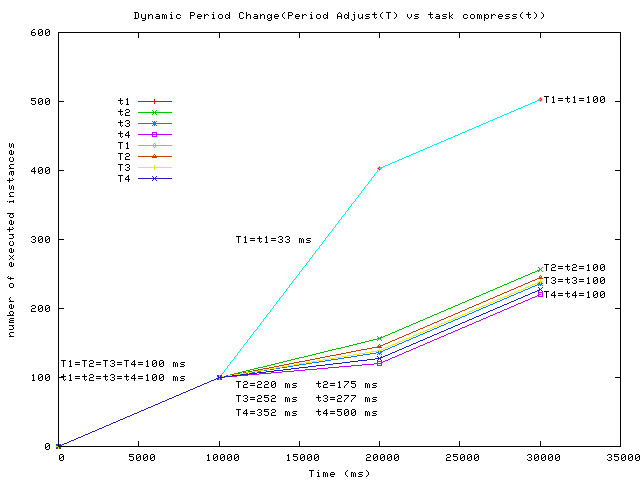}
\end{center}
\caption{Period\_Adjust vs Task\_compress }
\label{graph7}
\end{figure*}

\begin{figure*}[ht]
\begin{center}
\includegraphics[width=12cm, height=9cm]{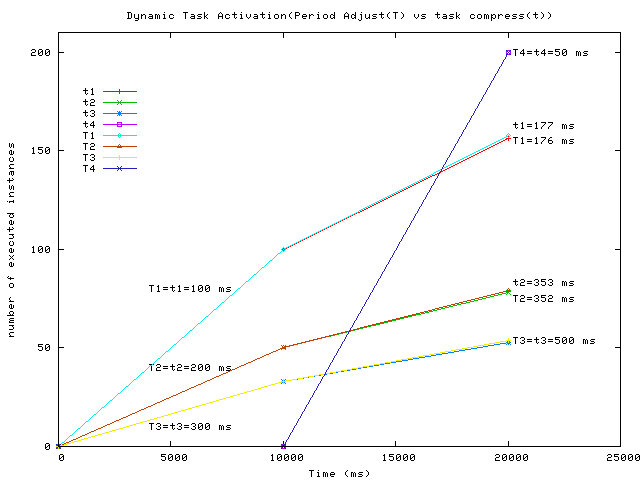}
\end{center}
\caption{Period\_Adjust vs Task\_compress }
\label{graph8}
\end{figure*}

\section{Conclusions and Future Work}
In this paper we have suggested Period\_Adjust algorithm for scheduling of tasks in which periods of soft real-time tasks are flexible. 
In this framework, periodic tasks can change their importance value to provide different quality of service. Importance value or weighting factor of soft real-time tasks are arranged in such a manner to keep the system underloaded. 
What makes Period\_Adjust more interesting is that it consider those soft real-time tasks whose periods are unbounded. The Period\_Adjust model is useful for supporting both multimedia systems and control applications in which the execution rates of some computational activities can not be properly predicted and they have to be dynamically tuned as a function of the current system state.\\
\indent We feel that Period\_Adjust model is a general model which can be applied in many applications. This framework can be extended to support the cases where deadline is less than period and computation time is variable. 

\end{document}